\documentclass[12pt]{article}
\usepackage{graphicx,setspace,fullpage, amsmath,bm, ulem} %
\usepackage{amsthm,amsmath,amsfonts,amssymb,bm,graphicx}
\usepackage{natbib}
\newtheorem{theorem}{Theorem}

\newtheorem{proposition}[theorem]{Proposition}

\title{On the Choice of Model Space Priors and Multiplicity Control in Bayesian Variable Selection: An Application to Streaming Logistic Regression}

\author{Joyee Ghosh\footnote{Joyee Ghosh is Associate Professor, Department of Statistics and Actuarial Science, The University of Iowa.}}

\date{}

\begin{document}

\maketitle

\begin{center}
 {Abstract}   
\end{center}
Bayesian variable selection (BVS) depends critically on the specification of a prior distribution over the model space, particularly for controlling sparsity and multiplicity. This paper examines the practical consequences of different model space priors for BVS in logistic regression, with an emphasis on streaming data settings. We review some popular and well-known Beta--Binomial priors alongside the recently proposed matryoshka doll (MD) prior. We introduce a simple approximation to the MD prior that yields independent inclusion indicators and is convenient for scalable inference. Using BIC-based approximations to marginal likelihoods, we compare the effect of different model space priors on posterior inclusion probabilities and coefficient estimation at intermediate and final stages of the data stream via simulation studies. Overall, the results indicate that no single model space prior uniformly dominates across scenarios, and that the recently proposed MD prior provides a useful additional option that occupies an intermediate position between commonly used Beta--Binomial priors with differing degrees of sparsity.

\doublespacing

\section{Introduction}
BVS provides a principled framework for incorporating variable selection uncertainty by placing a prior distribution over a space of candidate models, corresponding to different combinations of predictor variables in a regression set up. In regression problems with many predictors, the choice of the model space prior can substantially influence posterior inference, particularly posterior inclusion probabilities and selected model sizes.

A widely adopted choice is the Beta--Binomial prior (\cite{Scot:Berg:2010}), which is constructed by assigning a Beta prior to a common inclusion probability and assuming that, conditional on this probability, the predictor inclusion indicators are independent Bernoulli variables. One of the reasons for the popularity of this prior is better multiplicity control compared to the discrete uniform prior (\cite{Scot:Berg:2010}) when there are many predictors, and its associated computational convenience in posterior computation. Recently, the MD prior has been proposed by \cite{Woma:Tayl:Fuen:2025} as an alternative construction motivated by theoretical considerations of sparsity and multiplicity adjustment for large model spaces.

The outline of this paper is as follows. In Section 2, we provide a brief review of several model space priors that are commonly used or have been recently proposed in the BVS literature. In Section 3, we propose a simple approximation to the MD prior that preserves its limiting sparsity behavior with independent inclusion indicators, making it attractive for posterior inference in scenarios where independence or conditional independence has been proven to be crucial, such as Expectation Maximization algorithms with spike-and-slab priors (\cite{Rock:Geor:2014}, \cite{Rock:Geor:2018}). In Section 4, we compare these priors in the context of BVS for logistic regression with streaming data, focusing on posterior behavior at an intermediate batch and the final batch using BIC-based approximations to marginal likelihoods as in \cite{Ghos:Tan:Luo:2025}, \cite{De:Ghos:Ghos:2026}. Finally, in Section 5 we summarize our main findings and discuss some avenues for future research.

\section{Model Space Priors: A Brief Review}
Let $p$ denote the number of candidate predictors excluding the intercept, which is always included in the model and is not subject to selection. Let $\gamma ={(\gamma_1,\dots,\gamma_p)}^{\top}$ denote a vector of inclusion indicators, where $\gamma_j = 1$ indicates the inclusion of predictor $j$ and $\gamma_j = 0$ indicates the exclusion of predictor $j$. Let $k_{\gamma} = \sum_{j=1}^p \gamma_j$ denote the model size. Model space priors can be described by putting priors on the inclusion indicators or through the induced prior distribution on model size.

\subsection{Discrete Uniform Prior}
The discrete uniform prior assigns equal probability to all $2^p$ possible models, and is given by
\[
p(\gamma) = \frac{1}{2^{p}}.
\]
  Although simple and seemingly noninformative, this prior does not explicitly favor sparse models and can place substantial prior mass on large models even when $p$ is moderate (\cite{Clyd:Geor:2004}). It was shown by \cite{Scot:Berg:2010} 
  that this prior does not correct for multiplicity at all.

\subsection{Beta--Binomial Prior}
Under a Beta--Binomial prior, the inclusion indicators are conditionally independent given an inclusion probability $w$,
\[
\gamma_j \mid w \stackrel{iid}{\sim} {\rm Bernoulli}(w) \ {\rm for} \ j =1,2,\dots,p, \qquad w \sim {\rm Beta}(a,b), 
\]
where $a>0, b>0$. On integrating out $w$, this induces a Beta--Binomial distribution on model size (\cite{Wils:Iver:Clyd:Schm:Schi:2010}). Common choices include $\text{Beta}(1,1)$ (\cite{Scot:Berg:2010}), which corresponds to a uniform prior on $w$, as well as $\text{Beta}(1,p)$ and $\text{Beta}(1,p^2)$, which increasingly favor sparse models as $p$ grows. Alternative recommendations
have included $\text{Beta}(1,\sqrt{p})$
by \cite{Li:2013}, which was implemented by \cite{De:Ghos:2026} because they found that it led to improved posterior behavior under heavy-tailed error distributions, compared to $\text{Beta}(1,1)$ when $p$ was around a hundred. These priors are attractive due to their simplicity and ease of implementation, and they remain among the most commonly used choices in practice.

\subsection{Matryoshka Doll Prior}
The MD prior was recently proposed by \cite{Woma:Tayl:Fuen:2025} as a model space prior based on a simple and principled construction. Its defining feature is that, for any model, the prior odds of that model against the collection of all models that nest it is fixed at a constant value, denoted by $\xi$, and does not depend on the total number of available predictors. This construction provides an explicit form of multiplicity control, particularly suited to settings in which the true model is assumed to have finite size while the number of candidate predictors may grow. While this principle is theoretically appealing, it does not yield a closed-form expression for the model space prior. Instead, the MD prior probabilities can be computed recursively, which is feasible for moderately large values of $p$. For very large model spaces, direct computation becomes impractical, but the authors show that the prior can be accurately approximated using its limiting distribution as $p$ increases.

In particular, \cite{Woma:Tayl:Fuen:2025} showed that as $p \to \infty$, the model size distribution induced by the MD prior converges to a Poisson\((\theta)\) distribution, where $\theta=\log(1+(1/\xi))$. In particular,
\[
\lim_{p \to \infty} p\bigl(k_{\gamma} = k\bigr)
\;=\;
e^{-\theta}\frac{\theta^k}{k!}
\ {\rm for} \ k = 0,1,2,\ldots,
\]
and for finite $p$ the induced distribution can be approximated by a Poisson$(\theta)$ distribution truncated to $\{0,1,\dots,p\}$. From a practical standpoint, however, the MD prior induces dependence among inclusion indicators and cannot be naturally expressed as a collection of independent (or conditionally independent) Bernoulli variables. This dependence can complicate its implementation in algorithms in large-scale settings that benefit from conditional independence, such as EM algorithms with spike-and-slab priors or Gibbs samplers that exploit independence structure (\cite{Ghos:Clyd:2011}). This is the motivation for us to  consider an approximation to the MD prior that can be expressed in an independent Bernoulli form.

\section{A Bernoulli Approximation to the MD Prior}

In this section, we propose a simple and computationally convenient approximation to the MD prior with parameter $\theta>0$, obtained by assuming independent inclusion indicators. Specifically, we propose
\[
\gamma_j | \theta \stackrel{iid}{\sim} \mathrm{Bernoulli}(\theta/p) \ {\rm for} \ j=1,\ldots,p,
\]
which implies that $k_{\gamma} | \theta \sim \mathrm{Binomial}(p,\theta/p)$.

We show below that this approximation has some of the asymptotic properties of the MD prior with  parameter $\theta$, while differing in its finite $p$ properties.

\begin{proposition}
\label{prop1}
Fix $\theta>0$. Suppose $\gamma_j | \theta \stackrel{iid}{\sim} \mathrm{Bernoulli}(\theta/p)$ for $j=1,\ldots,p$, and 
$k_\gamma=\sum_{j=1}^p \gamma_j$. Then, for each fixed integer $k \ge 0$,
\[
\lim_{p\to\infty} p(k_{\gamma} = k | \theta)
=
e^{-\theta}\frac{\theta^k}{k!}.
\]
\end{proposition}
\noindent The proof is given in Appendix A.

\paragraph{Remark:}
Proposition \ref{prop1} shows that the asymptotic distribution of $k_{\gamma}$ is the same under the MD($\theta$) prior and 
the Bernoulli$(\theta/p)$ prior. Note that this is a special case of the standard Poisson approximation to a Binomial $(p,\theta/p)$ distribution.

\begin{proposition}
\label{prop2}
Fix $\theta>0$. Suppose $\gamma_j | \theta \stackrel{iid}{\sim} \mathrm{Bernoulli}(\theta/p)$ for $j=1,\ldots,p$, and let
$k_\gamma=\sum_{j=1}^p \gamma_j$. Then, for each fixed integer $k \ge 0$,
\[
\lim_{p\to\infty}
\frac{p(k_\gamma = k+1 | \theta)}{p(k_\gamma = k | \theta)}
=
\frac{\theta}{k+1}.
\]
\end{proposition}
\noindent The proof is provided in Appendix A.

\paragraph{Remark:}
 Note that the limiting value of the model size ratio in Proposition~\ref{prop2} coincides with that induced by the MD prior with parameter $\theta$, as shown by \cite{Woma:Tayl:Fuen:2025}. Here, model size refers to the number of predictors subject to variable selection that are included in the model; this matches the notion of model complexity in \cite{Woma:Tayl:Fuen:2025}, with our formulation corresponding to the special case in which only the intercept is treated as fixed, whereas their framework allows an arbitrary set of predictors to be always included. In particular, Proposition~7 of \cite{Woma:Tayl:Fuen:2025} establishes that under a Beta--Binomial prior obtained by placing a $\mathrm{Beta}(1,b)$ prior on the common inclusion probability, the limiting value of the same model size ratio as $p\to\infty$ equals (i) $1$ when $b$ is fixed, (ii) $1/(m+1)$ when $b=mp$, and (iii) $p^{\,1-v}$ when $b=p^v$, $v>1$.

\paragraph{Interpretation of Limiting Model Size Ratios:}
The above results provide insight about how different model space priors penalize increases in model size. 
The Beta$(1,1)$ prior on the common inclusion probability does not impose any additional penalty on expanding the model, as the prior odds of adding a predictor remain constant. The Beta$(1,p)$ prior induces a constant penalty on model expansion that depends on the parameter $m$, leading to uniform shrinkage toward smaller models. In contrast, the Beta$(1,p^2)$ prior yields a penalty that grows with $p$, which can result in aggressive shrinkage toward very sparse models. The MD$(\theta)$ prior and the Bernoulli$(\theta/p)$ approximation induce an adaptive penalty that depends on the current model size, leading to a gradual increase in resistance to model expansion as the model becomes larger.

While this Bernoulli approximation does not preserve the full structural properties of the MD prior, it yields independent inclusion indicators and is therefore particularly convenient for algorithms such as EM-type methods (\cite{Rock:Geor:2014}). In what follows, we assess how well this approximation mimics the behavior of the MD prior in practice.

\section{Bayesian Variable Selection for Streaming Logistic Regression}

In this section, we evaluate the model space priors described above in the context of
BVS for logistic regression with streaming data. We focus on
settings with $p=10$ and $p=15$ predictors (in addition to the intercept, which is always included), for which the
model space with $2^p$ models is small enough to be enumerated, allowing  Bayesian model averaging (BMA) without
Markov chain Monte Carlo (MCMC).

\subsection{Offline and Online Inference for Streaming Data}

We assume that observations arrive sequentially in batches
$D_1, D_2, \ldots, D_b, \dots$, where batch $b$ contains $n_b$ observations. Let
$D_b^\ast = \bigcup_{t=1}^b D_t$ denote the cumulative data up to batch $b$. We compare two approaches for posterior inference over the model space:

\begin{itemize}
\item \textbf{Offline} refers to recomputing posterior quantities at each batch using the
entire cumulative dataset $D_b^\ast$. This serves as a gold standard but requires storing
all past observations and refitting each candidate model as new data arrive.

\item \textbf{Online} refers to the Online~2 method of \cite{Ghos:Tan:Luo:2025}, which updates
posterior quantities sequentially using only the current batch data and a small set of summary
statistics from previous batches. This method is based on a second order Taylor expansion
of the log-likelihood and provides a more accurate approximation than the first-order
online method proposed in the same paper by \cite{Ghos:Tan:Luo:2025}. In the rest of this paper, Online refers exclusively to this
second-order method.
\end{itemize}

\subsection{Marginal Likelihood Approximation and Posterior Computation}

For a given model $\gamma = (\gamma_1,\ldots,\gamma_p)^{\top}$, let $X_\gamma$ denote the design
matrix containing the intercept and predictors with $\gamma_j=1$, and let $\beta_\gamma$
be the corresponding vector of regression coefficients. The logistic regression model is
\[
p(y_i=1 \mid x_{i,\gamma}, \gamma, \beta_\gamma)
=
\frac{\exp(x_{i,\gamma}^\top \beta_\gamma)}
     {1+\exp(x_{i,\gamma}^\top \beta_\gamma)},
\]
where $x_{i,\gamma}^\top$ denotes 
the $i$th row of $X_\gamma$.
Posterior model probabilities at batch $b$ are given by
\[
p(\gamma \mid D_b^\ast)
\;\propto\;
p(\gamma)\, m(D_b^\ast \mid \gamma),
\]
where $p(\gamma)$ is the model space prior and $m(D_b^\ast \mid \gamma)$ is the marginal
likelihood under model $\gamma$. To avoid specifying an explicit prior on $\beta_\gamma$ and to facilitate repeated batch
updates, we approximate the marginal likelihood using a BIC-based approximation as in \cite{Ghos:Tan:Luo:2025} as follows:
\[
\log m(D_b^\ast \mid \gamma)
\;\approx\;
\ell_b(\hat\beta_{\gamma,b}) - \frac{k_{\gamma}+1}{2}\log N_b,
\]
where $\ell_b(\hat\beta_{\gamma,b})$ is the log-likelihood evaluated at the maximum
likelihood estimate $\hat\beta_{\gamma,b}$ (or its online estimate under Online), $k_{\gamma}$ is the number of
included predictors (not including the intercept), and $N_b$ is the aggregate sample size at batch $b$.

Posterior inclusion probabilities are computed as
\[
p(\gamma_j=1 \mid D_b^\ast)
=
\sum_{\gamma:\,\gamma_j=1} p(\gamma \mid D_b^\ast),
\]
and BMA estimates of regression coefficients are obtained by taking weighted averages of
$\hat\beta_{\gamma,b}$ over models, with coefficients set to zero when excluded from a model, and the weights being equal to the posterior probabilities of models.

\subsection{Model Space Priors Under Comparison}

The following model space priors are compared, ordered from relatively weak to relatively strong sparsity-inducing property:

\begin{enumerate}
\item Discrete uniform prior on the model space.
\item Beta$(1,1)$ prior on the common inclusion probability.
\item Beta$(1,p)$ prior on the common inclusion probability.
\item MD prior with $\theta=1$.
\item Truncated Poisson approximation to the MD prior with $\theta=1$.
\item Bernoulli$(\theta/p)$ approximation proposed in Section~3 with $\theta=1$.
\item Beta$(1,p^2)$ prior on the common inclusion probability.
\end{enumerate}

The choice $\theta=1$ is adopted here, which is the default specification used by \cite{Woma:Tayl:Fuen:2025} and
corresponds to a Poisson$(1)$ limiting distribution for model size. This value provides a
natural baseline for sparsity and facilitates comparison between the MD prior and its
approximations.

\subsection{Simulation Scenarios}
We examine four simulation scenarios with two values of $p$ (10 and 15) and two sparsity regimes (sparse versus non-sparse). In all cases, covariates are
generated independently from  a normal distribution with mean 0 and variance 9. The intercept term is always included in the model. For each
scenario, we generate $25$ independent datasets.

\paragraph{Scenarios With $p=10$:}
These scenarios are implemented with $p=10$ covariates, other than the intercept. Batch
sizes are taken as $n_1=50$ and $n_b=10$ for $b=2,\ldots,21$, as the first batch size needs to be sufficiently large for the online method to work well (\cite{Luo:Song:2020, Ghos:Tan:Luo:2025}, \cite{De:Ghos:Ghos:2026}).
Let
$\beta =
( \beta_0,\beta_1,\ldots,\beta_{10} )^{\top}$, where $\beta_0$ is the intercept term. We have the following two cases:
\begin{itemize}
\item \textbf{Sparse:}  $\beta=(0.2,0.3,-0.4,0,\dots,0)^{\top}$.
\item \textbf{Non-sparse:} 
$\beta =(0.2,0.2,0.2,0.2,0.2,-0.2,-0.2,-0.2,0,0,0)^{\top}$.
\end{itemize}

\paragraph{Scenarios With $p=15$:}
Here we consider batch sizes $n_1=100$ and $n_b=10$ for $b=2,\ldots,21$ with the following two cases:
\begin{itemize}
\item \textbf{Sparse:}
$\beta = (0.2,\,0.3,\,0.4,\,0,\ldots,0)^{\top}$.

\item \textbf{Non-sparse:}
$\beta = (0.2,\,0.2,\,0.2,\,0.2,\,0.2,\,0.2,\,0.2,\,0.2,\,0,\ldots,0)^{\top}$.
\end{itemize}

\subsection{Performance Measures}
For each scenario and model space prior, we summarize the performance at an intermediate batch
($b=11$) and the final batch ($b=21$). Specifically, we report:
\begin{itemize}
\item Root mean squared error (RMSE) in estimating the true regression coefficient vector $\beta$ using BMA estimates of $\beta$.
\item RMSE in estimating the true model $\gamma$ using marginal posterior inclusion probabilities.
\end{itemize}
Box plots showing the values of RMSE across 25 replicates are used to summarize the variability for Offline and Online. We do not include running time comparisons in this study, as meaningful computational gains primarily arise for substantially larger batch sizes. In such regimes, however, posterior behavior under different model space priors becomes increasingly similar, making it difficult to isolate their individual effects.

\subsection{Summary of Simulation Results}
\begin{figure}[!h]
    \centering    \includegraphics[width=0.85\linewidth]{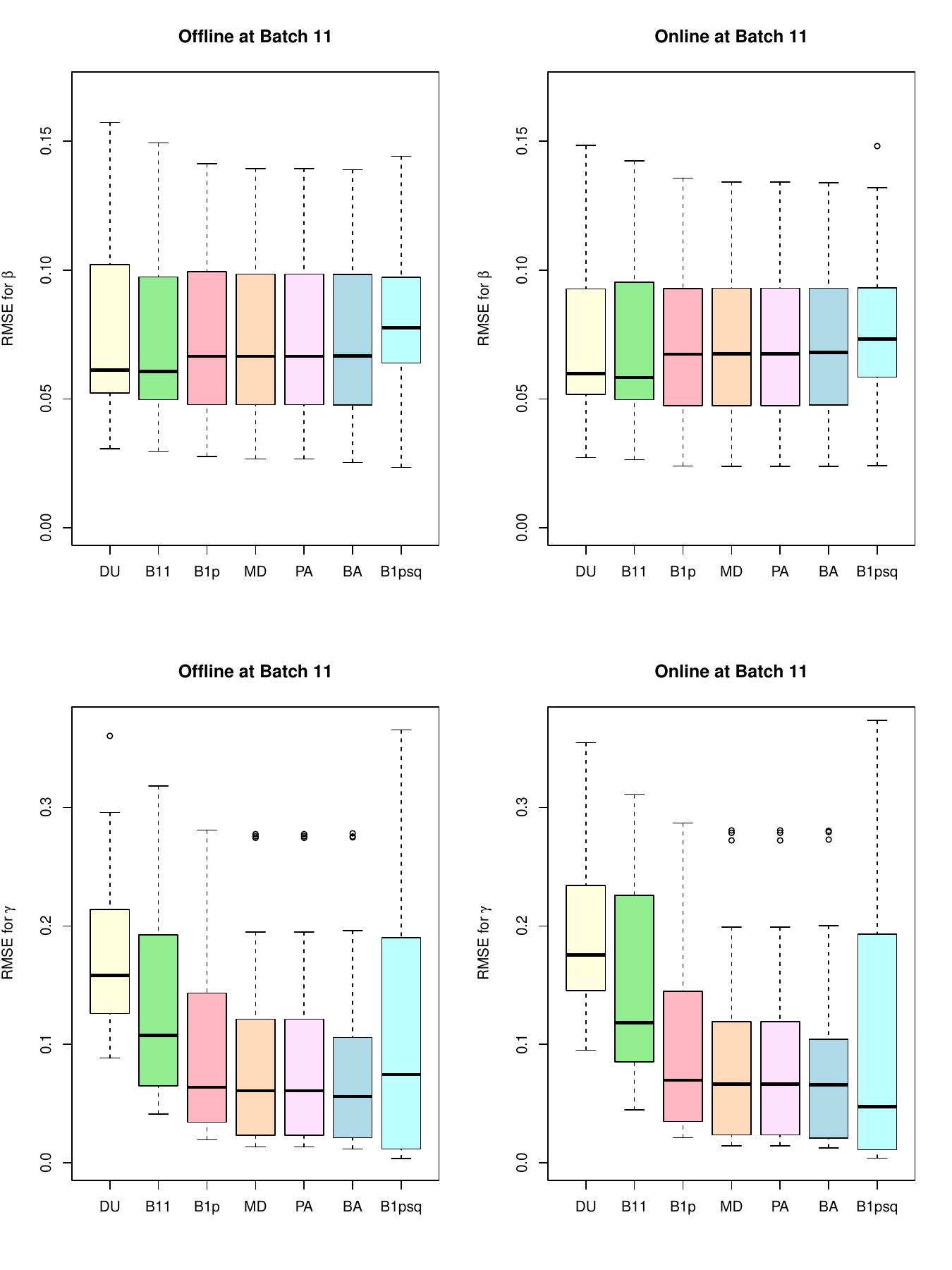}
    \caption{Box plots showing RMSE for estimating $\beta$ and $\gamma$ at batch 11, across 25 replicates for Offline and Online methods, for \uline{\textit{sparse simulation scenario with $p=10$}} under 7 model space priors (DU: discrete uniform, B11: Beta($1,1$), B1p: Beta($1,p$), MD: matryoshka doll, PA: Poisson (truncated) approximation to MD, BA: Bernoulli approximation to MD,  B1psq: Beta($1,p^2$)).}
    \label{fig1}
\end{figure}

\begin{figure}[!h]
    \centering    \includegraphics[width=0.85\linewidth]{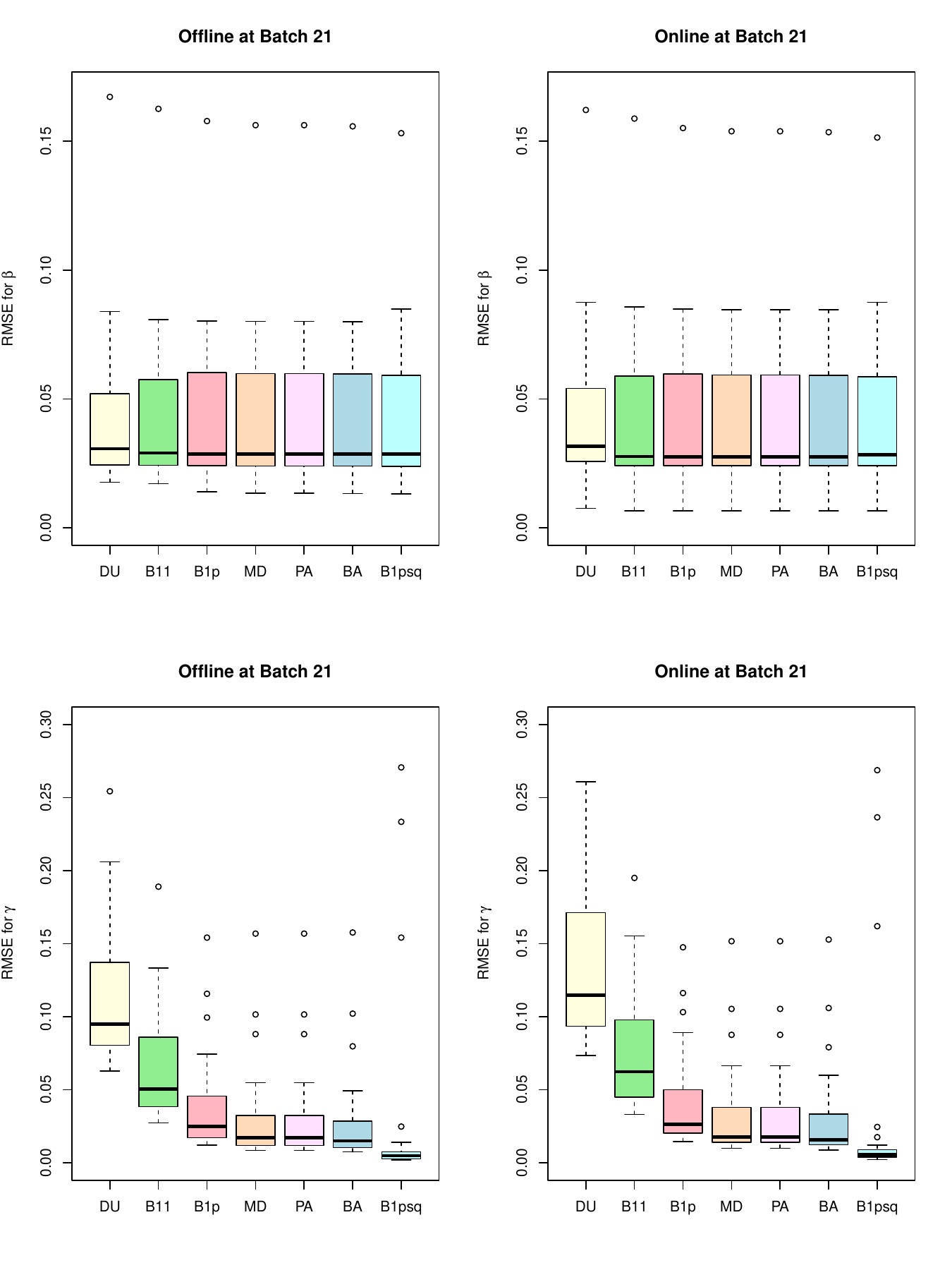}
    \caption{Box plots showing RMSE for estimating $\beta$ and $\gamma$ at batch 21, across 25 replicates for Offline and Online methods, for \uline{\textit{sparse simulation scenario with $p=10$}} under 7 model space priors (DU: discrete uniform, B11: Beta($1,1$), 
    B1p: Beta($1,p$), MD: matryoshka doll, PA: Poisson (truncated) approximation to MD, BA: Bernoulli approximation to MD,  
    B1psq: Beta($1,p^2$)).}
    \label{fig2}
\end{figure}

\begin{figure}[!h]
    \centering    \includegraphics[width=0.85\linewidth]{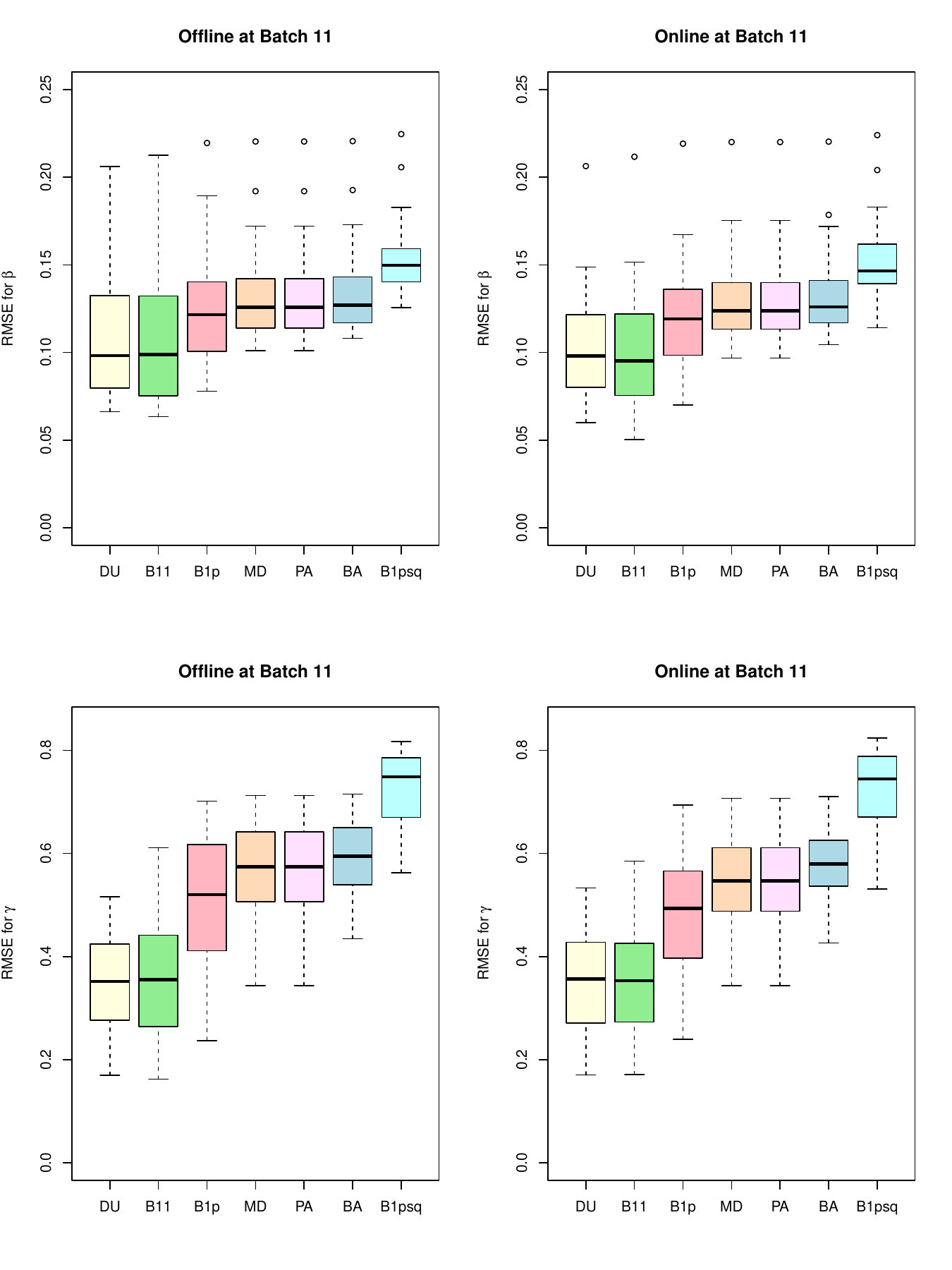}
    \caption{Box plots showing RMSE for estimating $\beta$ and $\gamma$ at batch 11, across 25 replicates for Offline and Online methods, for \uline{\textit{non-sparse simulation scenario with $p=10$}} under 7 model space priors (DU: discrete uniform, B11: Beta($1,1$), B1p: Beta($1,p$), MD: matryoshka doll, PA: Poisson (truncated) approximation to MD, BA: Bernoulli approximation to MD, B1psq: Beta($1,p^2$)).}
    \label{fig3}
\end{figure}

\begin{figure}[!h]
    \centering    \includegraphics[width=0.85\linewidth]{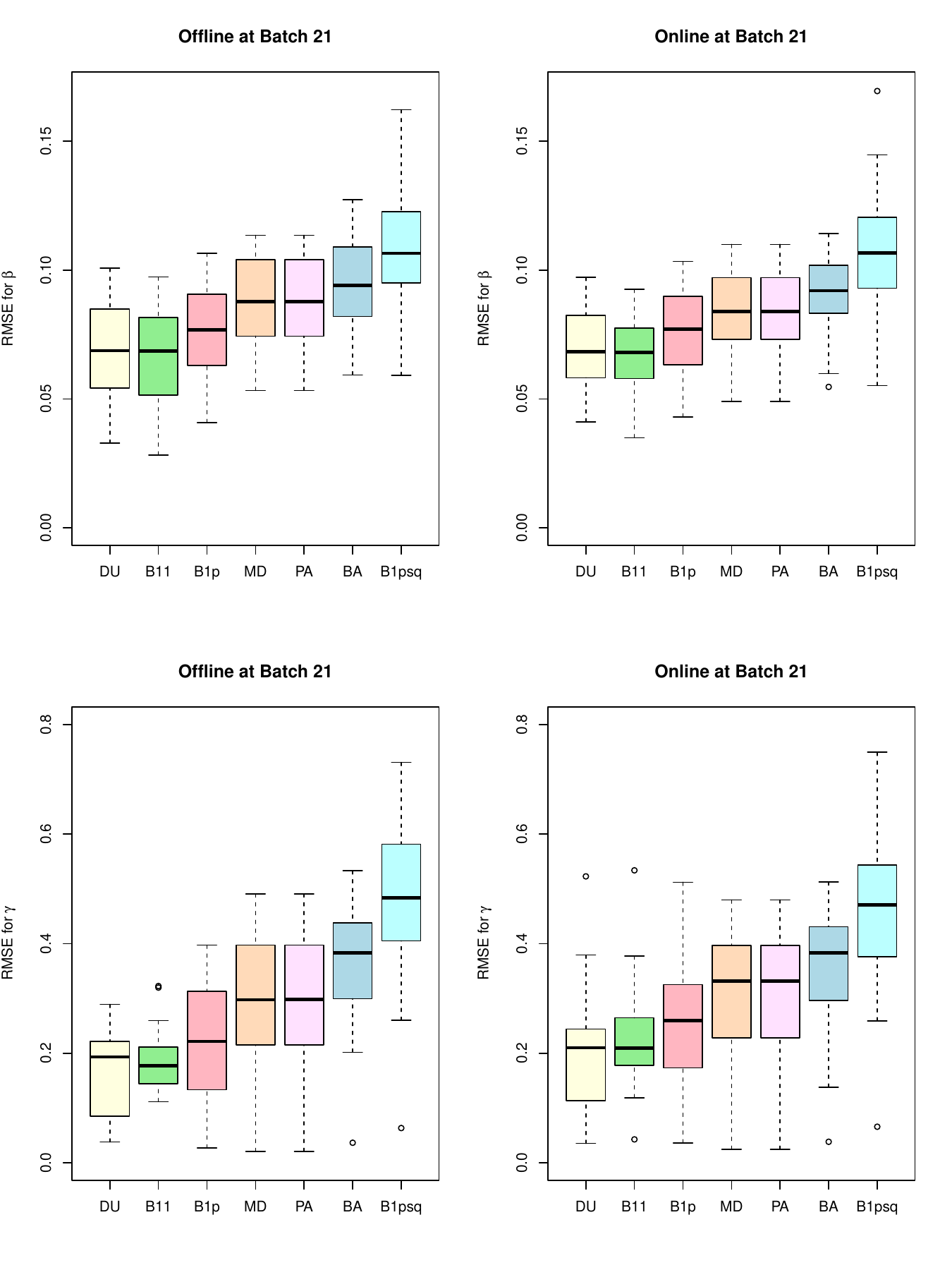}
    \caption{Box plots showing RMSE for estimating $\beta$ and $\gamma$ at batch 21, across 25 replicates for Offline and Online methods, for \uline{\textit{non-sparse simulation scenario with $p=10$}} under 7 model space priors (DU: discrete uniform, B11: Beta($1,1$), B1p: Beta($1,p$), MD: matryoshka doll, PA: Poisson (truncated) approximation to MD, BA: Bernoulli approximation to MD, B1psq: Beta($1,p^2$)).}
    \label{fig4}
\end{figure}

\begin{figure}[!h]
    \centering    \includegraphics[width=0.85\linewidth]{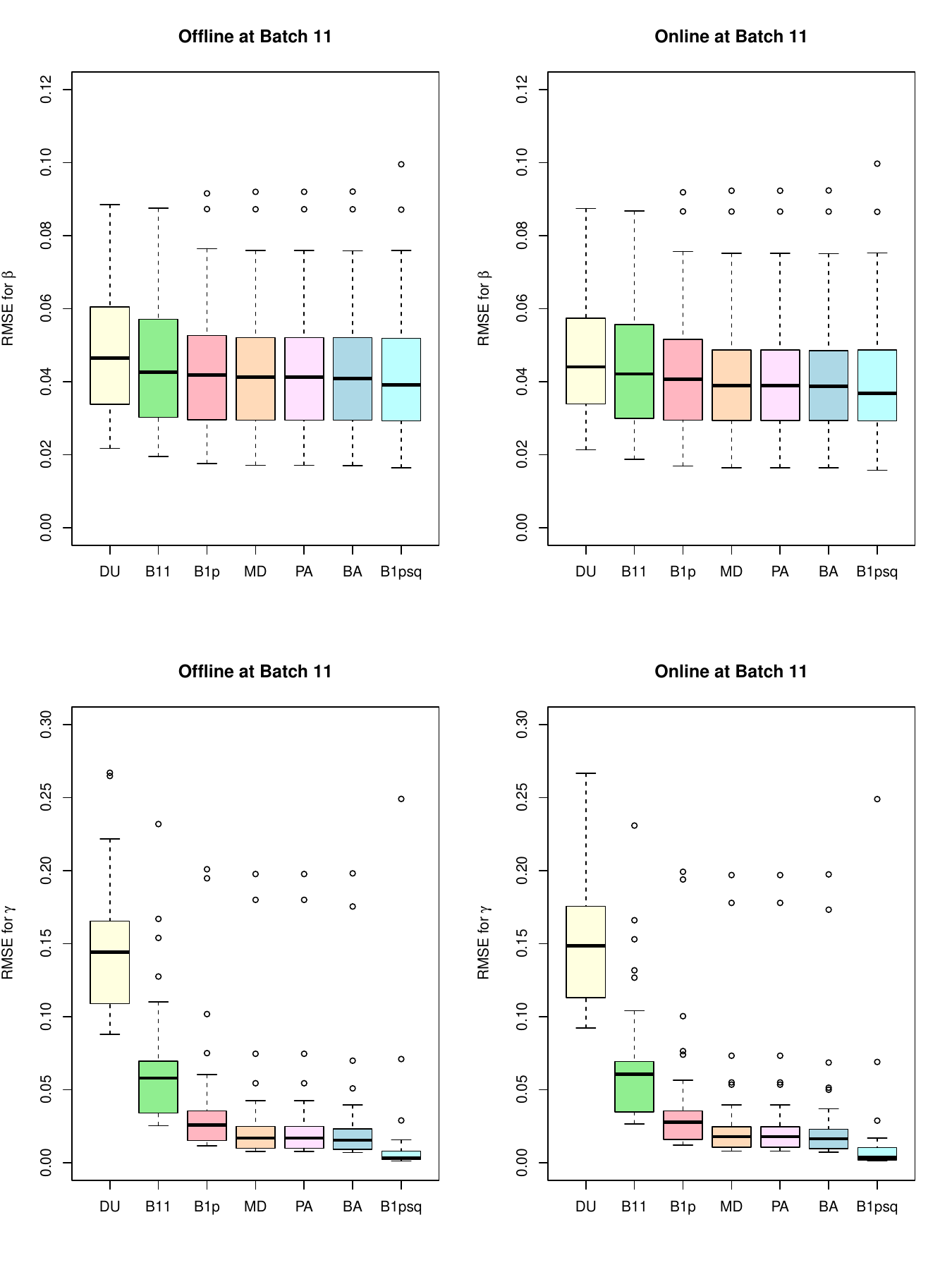}
    \caption{Box plots showing RMSE for estimating $\beta$ and $\gamma$ at batch 11, across 25 replicates for Offline and Online methods, for \uline{\textit{sparse simulation scenario with $p=15$}} under 7 model space priors (DU: discrete uniform, B11: Beta($1,1$), B1p: Beta($1,p$), MD: matryoshka doll, PA: Poisson (truncated) approximation to MD, BA: Bernoulli approximation to MD, B1psq: Beta($1,p^2$)).}
    \label{fig5}
\end{figure}

\begin{figure}[!h]
    \centering    \includegraphics[width=0.85\linewidth]{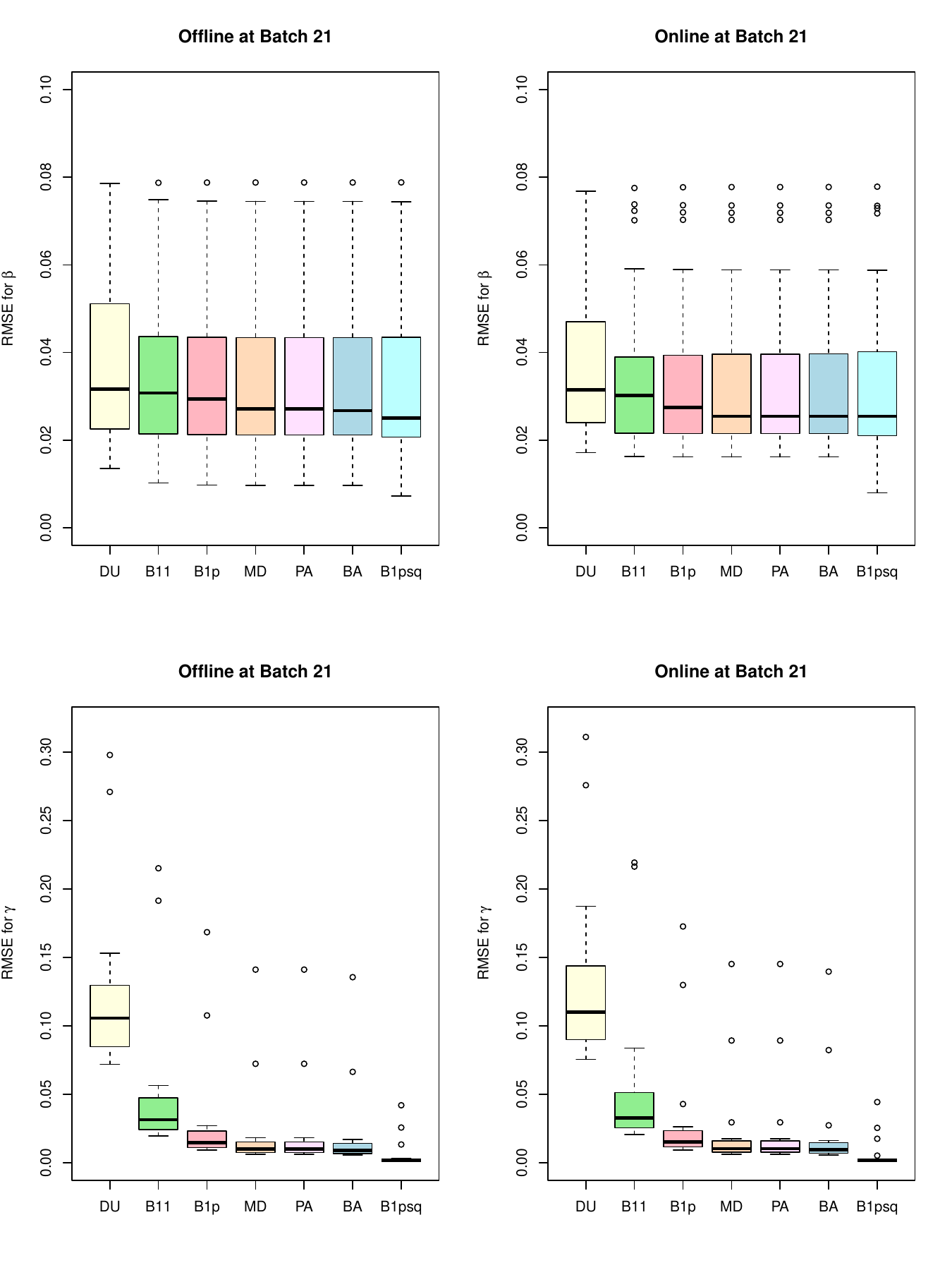}
    \caption{Box plots showing RMSE for estimating $\beta$ and $\gamma$ at batch 21, across 25 replicates for Offline and Online methods, for \uline{\textit{sparse simulation scenario with $p=15$}} under 7 model space priors (DU: discrete uniform, B11: Beta($1,1$), B1p: Beta($1,p$), MD: matryoshka doll, PA: Poisson (truncated) approximation to MD, BA: Bernoulli approximation to MD, B1psq: Beta($1,p^2$)).}
    \label{fig6}
\end{figure}

\begin{figure}[!h]
    \centering    \includegraphics[width=0.85\linewidth]{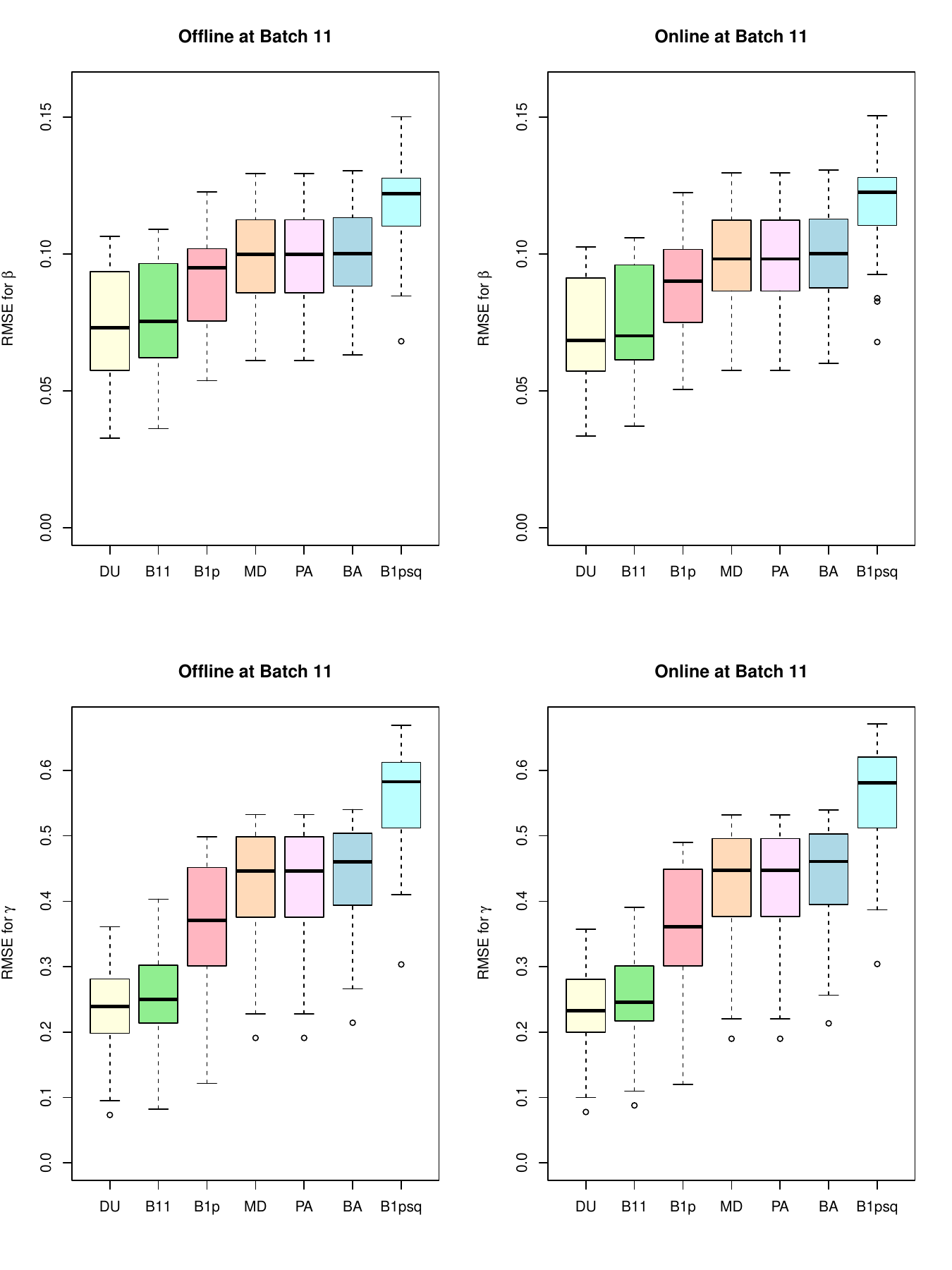}
    \caption{Box plots showing RMSE for estimating $\beta$ and $\gamma$ at batch 11, across 25 replicates for Offline and Online methods, for \uline{\textit{non-sparse simulation scenario with $p=15$}} under 7 model space priors (DU: discrete uniform, B11: Beta($1,1$), B1p: Beta($1,p$), MD: matryoshka doll, PA: Poisson (truncated) approximation to MD, BA: Bernoulli approximation to MD, B1psq: Beta($1,p^2$)).}
    \label{fig7}
\end{figure}

\begin{figure}[!h]
    \centering    \includegraphics[width=0.85\linewidth]{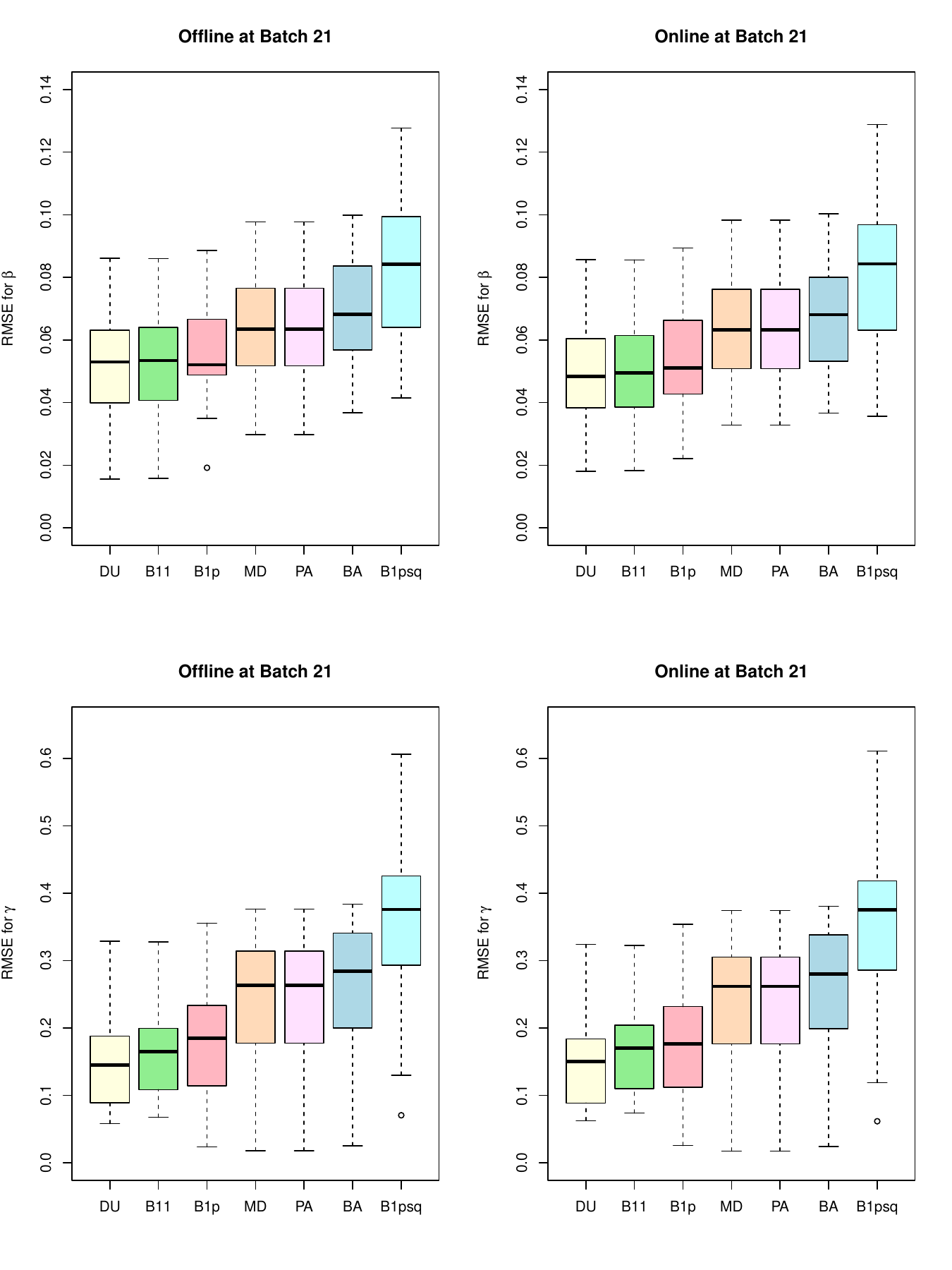}
    \caption{Box plots showing RMSE for estimating $\beta$ and $\gamma$ at batch 21, across 25 replicates for Offline and Online methods, for \uline{\textit{non-sparse simulation scenario with $p=15$}} under 7 model space priors (DU: discrete uniform, B11: Beta($1,1$), B1p: Beta($1,p$), MD: matryoshka doll, PA: Poisson (truncated) approximation to MD, BA: Bernoulli approximation to MD, B1psq: Beta($1,p^2$)).}
    \label{fig8}
\end{figure}

The box plots for the four simulation scenarios at two batches ($b=11$ and $b=21$) under each scenario are provided in Figures~\ref{fig1}--\ref{fig8}. We provide an overall summary of the results here. Across all scenarios, Offline and Online produce very similar posterior behavior, with Online closely tracking Offline even at intermediate batches. This agreement persists across different model space priors and both sparse and non-sparse regimes, indicating that the online procedure provides an accurate approximation to offline inference in the settings considered here. Differences among model space priors are evident in both sparse and non-sparse settings, though their nature differs substantially across regimes. 

In sparse scenarios, priors that more strongly favor sparse models, such as the Beta$(1,p^2)$ prior and, to a smaller extent, the MD prior and its two approximations, tend to achieve lower RMSE for posterior inclusion probabilities by more effectively controlling false positives. In contrast, priors that induce little or no multiplicity control, such as the discrete uniform prior and the Beta$(1,1)$ prior, exhibit higher RMSE for inclusion probabilities due to the inclusion of additional noise predictors. The Beta$(1,p)$ prior typically falls between these two extremes, reflecting its intermediate level of sparsity promotion.

In non-sparse scenarios, this pattern is largely reversed. Priors that strongly favor small models tend to incur higher RMSE for inclusion probabilities due to the exclusion of truly active predictors, resulting in false negatives. In these settings, less sparsity-inducing priors, most notably the discrete uniform and Beta$(1,1)$ priors, perform better in terms of inclusion probability RMSE, while the Beta$(1,p^2)$ prior performs the worst. The MD prior and its approximations again exhibit intermediate behavior, neither consistently dominating nor performing poorly across all cases.

Across all scenarios, differences among model space priors have a much larger impact on the estimation of posterior inclusion probabilities than on the estimation of regression coefficients. In the sparse settings, even when false positives occur under less sparsity-promoting priors, the corresponding regression coefficients are typically shrunk toward zero, leading to relatively modest differences in RMSE for coefficient estimation. In contrast, in the non-sparse scenarios, differences in coefficient RMSE become more pronounced: priors that aggressively favor sparsity tend to drop truly nonzero coefficients, resulting in higher RMSE due to false negatives, whereas priors that allow larger models achieve better coefficient recovery.

Finally, the two approximations to the MD prior: (1) the truncated Poisson approximation proposed by \cite{Woma:Tayl:Fuen:2025}, and (2) the Bernoulli$(\theta/p)$ approximation proposed by us in this paper, track the behavior of the MD prior reasonably closely across all scenarios considered, with the posterior under approximation (1) being almost indistinguishable from the posterior under the MD prior, whereas (2) shows a small difference. However, differences among these three priors are generally small relative to differences between broader classes of model space priors, suggesting that the proposed Bernoulli$(\theta/p)$ approximation provides a reasonable and computationally convenient surrogate for the MD prior in finite-$p$ settings, where independence among inclusion indicators may be vital for computational reasons in an EM algorithm.

\section{Discussion}
In this paper, we examined the role of the model space prior in BVS for logistic regression, with a particular emphasis on streaming data settings. We reviewed several commonly used priors, including the discrete uniform and Beta--Binomial families, alongside the recently proposed MD prior (\cite{Woma:Tayl:Fuen:2025}). We also introduced a simple Bernoulli$(\theta/p)$ approximation to the MD prior that preserves its limiting sparsity behavior while yielding independent inclusion indicators, which is attractive for scalable inference in EM-type algorithms.

From a theoretical perspective, the MD prior provides an appealing framework for multiplicity control, inducing Poisson-type behavior in model size and an adaptive penalty for adding an additional predictor that depends on the current model size rather than directly on the total number of predictors. In contrast, for adding a predictor to a current model, Beta--Binomial priors impose penalties whose strength depends on how the Beta hyperparameters change with 
$p$, ranging from no additional penalty under the Beta$(1,1)$ prior
to very aggressive penalty under the Beta($1,p^2$) prior. The Beta($1,p$)
 prior occupies an intermediate position, imposing a constant penalty on the prior odds of adding a predictor. Our results suggest that in finite-$p$ regimes with relatively modest values of $p$, the practical behavior of the MD prior is often comparable to that of the Beta$(1,p)$ prior, though the two arise from different conceptual motivations.

Our empirical results under logistic regression and streaming data illustrate that no single model space prior uniformly dominates across scenarios. In sparse settings, stronger sparsity-inducing priors tend to better control false positives, whereas, in non-sparse settings, milder priors often perform better by avoiding false negatives. The MD prior and its approximations generally exhibit behavior between these extremes, consistent with their theoretical construction. The two approximations to the MD prior track its behavior closely in finite-$p$ regimes, with the truncated Poisson approximation being nearly indistinguishable from the MD prior and the Bernoulli$(\theta/p)$ approximation showing only minor deviations while offering substantial computational convenience.

The settings considered here involve relatively small model spaces that allow the enumeration of all candidate models with BIC-based marginal likelihood approximations. An important direction for future work is to investigate the behavior of the MD prior and its approximations in settings with larger values of $p$, where full model space enumeration is computationally infeasible and posterior inference relies on MCMC or other stochastic search methods (\cite{De:Ghos:Ghos:2026}). Understanding how the theoretical advantages of the MD prior translate to such high-dimensional regimes remains an interesting and open problem.

\bibliography{md.bib}

\section*{Appendix A}

\noindent \textbf{Proposition \ref{prop1}:} 
Fix $\theta>0$. Suppose $\gamma_j | \theta \stackrel{iid}{\sim} \mathrm{Bernoulli}(\theta/p)$ for $j=1,\ldots,p$, and 
$k_\gamma=\sum_{j=1}^p \gamma_j$. Then, for each fixed integer $k \ge 0$,
\[
\lim_{p\to\infty} p(k_{\gamma} = k | \theta)
=
e^{-\theta}\frac{\theta^k}{k!}.
\]

\begin{proof}
For each $p$, $k_\gamma | \theta \sim \mathrm{Binomial}(p,\theta/p)$, hence for fixed $k$,
\begin{equation}
p(k_{\gamma} = k | \theta)
=
\binom{p}{k}\left(\frac{\theta}{p}\right)^k\left(1 - \frac{\theta}{p}\right)^{p-k}.
\label{eq1}
\end{equation}

\noindent We simplify the product of the first two terms in the right hand side (RHS) of (\ref{eq1}) below.


\[
\binom{p}{k}\left(\frac{\theta}{p}\right)^k
=
\frac{\theta^k}{k!}\cdot \frac{p(p-1)\cdots(p-k+1)}{p^k}
=
\frac{\theta^k}{k!}\prod_{j=0}^{k-1}\left(1-\frac{j}{p}\right).
\]
Since $k$ is fixed, each factor in the above product term converges to $1$ as $p\to\infty$, and we have
\begin{equation}
    \lim_{p\to\infty} \binom{p}{k}\left(\frac{\theta}{p}\right)^k = \frac{\theta^k}{k!}.
\label{eq2}
\end{equation}

\noindent We now simplify the third term in the RHS of (\ref{eq1}) below.

\[
\left(1-\frac{\theta}{p}\right)^{p-k}
=
\left(1-\frac{\theta}{p}\right)^p
\left(1-\frac{\theta}{p}\right)^{-k}.
\]
Because $\theta$ is fixed, the standard limit $(1-\theta/p)^p \to e^{-\theta}$ holds as $p\to\infty$. Moreover, since $k$ is fixed,
\[
\lim_{p\to\infty}  \left(1-\frac{\theta}{p}\right)^{-k} = 1.
\]
Combining these gives
\begin{equation}
  \lim_{p\to\infty} \left(1-\frac{\theta}{p}\right)^{p-k} = e^{-\theta}.  
  \label{eq3}
\end{equation}

\noindent Combining (\ref{eq2}) and (\ref{eq3}) proves the claim.
\end{proof}

\noindent \textbf{Proposition \ref{prop2}:}
Fix $\theta>0$. Suppose $\gamma_j | \theta \stackrel{iid}{\sim} \mathrm{Bernoulli}(\theta/p)$ for $j=1,\ldots,p$, and let
$k_\gamma=\sum_{j=1}^p \gamma_j$. Then, for each fixed integer $k \ge 0$,
\[
\lim_{p\to\infty}
\frac{p(k_\gamma = k+1 | \theta)}{p(k_\gamma = k | \theta)}
=
\frac{\theta}{k+1}.
\]

\begin{proof}
Since $k_\gamma | \theta \sim \mathrm{Binomial}(p,\theta/p)$,
\[
p(k_\gamma=k |\theta)
=
\binom{p}{k}\left(\frac{\theta}{p}\right)^k\left(1-\frac{\theta}{p}\right)^{p-k}.
\]
Therefore, for fixed $k\ge 0$,
\begin{align*}
\frac{p(k_\gamma = k+1 | \theta)}{p(k_\gamma = k | \theta)}
&=
\frac{\binom{p}{k+1}\left(\frac{\theta}{p}\right)^{k+1}\left(1-\frac{\theta}{p}\right)^{p-k-1}}
{\binom{p}{k}\left(\frac{\theta}{p}\right)^k\left(1-\frac{\theta}{p}\right)^{p-k}} \\
&=
\frac{\binom{p}{k+1}}{\binom{p}{k}}
\cdot \frac{\theta}{p}
\cdot \left(1-\frac{\theta}{p}\right)^{-1} \\
&=\frac{p-k}{k+1}\cdot \frac{\theta}{p}\cdot \frac{p}{p-\theta} \\
&= \frac{\theta}{k+1}\cdot \frac{p-k}{p-\theta}.
\end{align*}
Because $\theta$ and $k$ are fixed,
$(p-k)/(p-\theta) \rightarrow 1$ as
$p \rightarrow \infty$, 
which proves the claim.
\end{proof}

\end{document}